\documentclass[aps,pra,twocolumn,superscriptaddress,showpacs]{revtex4}

\usepackage{CJK}

\usepackage{graphicx}
\newcommand{\tr}{\mbox{Tr} }
\newcommand{\ket}[1]{\left | #1 \right \rangle}

\begin{document}
\title{Joint measurements and Svetlichny's inequality}

\author{Yang Xiang}
\email{njuxy@sohu.com}
\affiliation{School of Physics and Electronics, Henan University, Kaifeng, Henan 475004, China}

\author{Wei Ren}
\email{weiren@uark.edu}
\affiliation{Physics Department, University of Arkansas, Fayetteville, Arkansas 72701, USA}
\date{\today}
\begin{abstract}
We prove that the Svetlichny's inequality can be derived from the existence of joint measurements and the principle of no-signaling. Then we show that, on the basis of quantum measurement assumption, it would imply the breach of causality if the magnitude of violation of Svetlichny's inequality exceeds quantum bound.

\end{abstract}

\pacs{03.65.Ud, 03.65.Ta, 03.67.-a}
\maketitle




\section{introduction}

Heisenberg uncertainty \cite{hei} forbids in principle that non-commuting observables can be simultaneously measured with arbitrary accuracy.
The extreme opposite of Hersenberg uncertainty is the Einstein's reality \cite{ein}, and Bell's inequality \cite{bell} originally derived just from
the assumption of the Einstein's reality and the no-signaling principle. However, quantum mechanics allows unsharp simultaneous measurements of two
non-commuting observables \cite{art1,art2,sten}. This means that if we tolerate an increase in the variance of these two non-commuting
observables we can perform a joint measurement of them. Here, joint measurement is defined so that one measurement on a single physical system
simultaneously produces values for more than one observables. If these two observables mutually commute, it is obvious that
the joint measurements of them can be accomplished with projective quantum measurements; if these two observables do not mutually commute,
in order to carry out the joint measurement one has to adopt positive operator valued measures (POVMs) \cite{and1,and2,busch}. Essentially,
the existence of joint measurements of some observables demands that a joint probability distribution of the values of these observables exists.
Back in 1982, Fine \cite{fine} has pointed out that the existence of a joint probability distribution for the values of the observables which
involved in a Bell's inequality must result in the satisfaction of the Bell's inequality. Along this line of thought, Andersson \textit{et al.}
\cite{and1,and2} replaced the Einstein's reality with the joint measurements and showed that from joint measurements and no-signaling principle
Bell's inequality can also be derived. This indicates that the results of joint measurements result in the satisfaction of Bell's inequality.
They also found that the Bell's inequality naturally provides a tight bound on the sharpness for the joint measurements, and this bound was
first derived by Busch \cite{bus1}.

The purpose of this work is to introduce joint measurements into the study of genuine multipartite correlations that cannot be reduced to mixtures of states
in which a smaller number of subsystems are entangled, and to investigate the connection between them.
The violation of Svetlichny's inequality (SI) \cite{svet} is a confirmation of genuine multipartite correlations, and the magnitude of violation of SI is usually
regarded as a measure of it. We show that, if the no-signaling principle is available and joint measurements are made on one particle of a $N$-particle quantum system,
then the $N$-particle SI can be derived. This means that the existence of joint measurements must produce the absence of genuine multipartite correlations.
Finally, we give a discussion about the quantum bound on the violation of SI.

\section{derivation of Svetlichny's inequality}

In this section, we only give the derivation of SI by assuming the existences of joint measurements and no-signaling principle,
and present how to achieve joint measurements in the next section.

{\it Derivation of three-particle SI}~~Three-particle SI \cite{svet} can distinguish
between genuine three-particle correlations and
two-particle correlations. This means that one can
find a violation of SI inequality if and only if there exist genuine three-particle correlations in a three-particle setting.
Consider three observers, Alice, Bob, and Carol, who share three entangled qubits. Each of them can choose to measure one
of two dichotomous observables. We denote $A$ and $A'$ as Alice's measurement results when she performs measurement $a$ and $a'$ respectively, and
similarly $B$, $B'$, $b$, and $b'$ ($C$, $C'$, $c$, and $c'$) for Bob's (Carol's), and the measurement results of all observables can be $-1$ or $+1$. Then SI is expressed as \cite{svet}
\begin{eqnarray}
S_{3}&\equiv&|E(ABC)+E(ABC')+E(A B'C)\nonumber\\
&&+E(A'BC)-E(A B'C')-E(A'BC')\nonumber\\
&&-E(A'B'C)-E(A'B'C')|\leq 4,
\label{si3}
\end{eqnarray}
where $S_{3}$ is the three-particle Svetlichny's operator, and $E(ABC)$ represents the expectation value of the product of the measurement outcomes of the observables $a$, $b$, and $c$.
It was shown by Svetlichny \cite{svet} that quantum predictions can violate his inequality, and the maximum violation ($S=4\sqrt{2}$)
allowed in quantum mechanics can be achieved with GHZ states \cite{mit}.

Now we assume that Alice performs a joint measurement of $a$ and $a'$, and denote the measurement results as $A_{J}$ and $A_{J}^{'}$.
We can get
\begin{eqnarray}
&&E(A_{J}BC)+E(A_{J}^{'}BC)\nonumber\\
&=&P(A_{J}=A_{J}^{'}=BC)+P(A_{J}=-A_{J}^{'}=BC)\nonumber\\
&&-P(A_{J}=A_{J}^{'}=-BC)-P(A_{J}=-A_{J}^{'}=-BC)\nonumber\\
&&+P(A_{J}^{'}=A_{J}=BC)+P(A_{J}^{'}=-A_{J}=BC)\nonumber\\
&&-P(A_{J}^{'}=A_{J}=-BC)-P(A_{J}^{'}=-A_{J}=-BC)\nonumber\\
&=&2\big[P(A_{J}=A_{J}^{'}=BC)-P(A_{J}^{'}=A_{J}=-BC)\big]\nonumber\\
&\leq& 2\big[P(A_{J}=A_{J}^{'}=BC)+P(A_{J}^{'}=A_{J}=-BC)\big]\nonumber\\
&=&2P(A_{J}=A_{J}^{'};bc),
\label{d1}
\end{eqnarray}
where $P$ is probability function, and $P(A_{J}=A_{J}^{'};bc)$ represents the probability of that Alice obtains $A_{J}=A_{J}^{'}$ when
Bob and Carol respectively choose performing measurement $b$ and $c$.

Similarly, we can get
\begin{eqnarray}
&&E(A_{J}BC')-E(A_{J}^{'}BC')\nonumber\\
&=&2\big[P(A_{J}=-A_{J}^{'}=BC')-P(-A_{J}^{'}=A_{J}=-BC')\big]\nonumber\\
&\leq&2\big[P(A_{J}=-A_{J}^{'}=BC')+P(-A_{J}^{'}=A_{J}=-BC')\big]\nonumber\\
&=&2P(A_{J}=-A_{J}^{'};bc'),
\label{d2}
\end{eqnarray}
\begin{eqnarray}
E(A_{J}B'C)-E(A_{J}^{'}B'C)\leq2P(A_{J}=-A_{J}^{'};b'c),
\label{d3}
\end{eqnarray}
and
\begin{eqnarray}
E(A_{J}B'C')+E(A_{J}^{'}B'C')\leq2P(A_{J}=A_{J}^{'};b'c').
\label{d4}
\end{eqnarray}

From Eq. (\ref{si3}), Eq. (\ref{d1}), Eq. (\ref{d2}), Eq. (\ref{d3}), and Eq. (\ref{d4}), we obtain
\begin{eqnarray}
S_{3}^{J}&\leq& 2P(A_{J}=A_{J}^{'};bc)+2P(A_{J}=A_{J}^{'};b'c')\nonumber\\
&&+2P(A_{J}=-A_{J}^{'};bc')+2P(A_{J}=-A_{J}^{'};b'c),\nonumber\\
\label{si33}
\end{eqnarray}
where we denote $S_{3}^{J}$ as the three-particle Svetlichny's operator concerned with joint measurements.
Due to the no-signaling principle, the probability of Alice getting $A_{J}=A_{J}^{'}$ or $A_{J}=-A_{J}^{'}$
should be independent of the measurement choices of Bob and Carol, i.e.
\begin{eqnarray}
&&P(A_{J}=A_{J}^{'};bc)=P(A_{J}=A_{J}^{'};b'c')=P(A_{J}=A_{J}^{'}),\nonumber\\
&&P(A_{J}=-A_{J}^{'};bc')=P(A_{J}=-A_{J}^{'};b'c)\nonumber\\
&&=P(A_{J}=-A_{J}^{'})
\label{d5}
\end{eqnarray}
From Eq. (\ref{si33}) and Eq. (\ref{d5}) we finally get $S_{3}^{J}\leq 4$, i.e. the three-particle SI.

{\it Derivation of $N$-particle SI}
~~Suppose there are $N$ players who share $N$ particles, each one of them performs dichotomous measurements on each of the $N$ particles.
The measurement settings are represented by $x_{1}$, $x_{2}$,...$x_{N}$, respectively, with possible values $0,1$. The measurement
results are represented by $A_{1}$, $A_{2}$,...$A_{N}$, respectively, and with possible values $-1,1$.
Then the $N$-particle SI can be expressed as \cite{seev}
\begin{eqnarray}
S_{N}&\equiv&\left|\sum_{\{x_{i}\}} {v(x_{1},x_{2},...,x_{N})E(A_{1}A_{2}\cdot\cdot\cdot A_{N}|x_{1},x_{2},...,x_{N})}\right|\nonumber\\
&\leq& 2^{N-1},
\label{sin}
\end{eqnarray}
where $S_{N}$ is the $N$-particle Svetlichny's operator, $\{x_{i}\}$ stands for an $N$-tuple $x_{1},...,x_{N}$, $E(A_{1}A_{2}\cdot\cdot\cdot A_{N}|x_{1},x_{2},...,x_{N})$ represents the
expectation value of the product of the measurement
outcomes of the observables $x_{1},x_{2},...,x_{N}$,
and $v(x_{1},x_{2},...,x_{N})$ is a sign function given by
\begin{eqnarray}
v(x_{1},x_{2},...,x_{N})=(-1)^{[\frac{k(k-1)}{2}]},
\end{eqnarray}
where $k$ is the number of times index $1$ appears in $(x_{1},x_{2},...,x_{N})$.

Without losing generality, we assume the first player makes a joint measurement of $x_{1}=0$ and $x_{1}=1$ on the first particle, with
results $A_{1}^{J}$ and $A_{1}^{J'}$. We note that the summation in Eq. (\ref{sin}) can be expressed as
\begin{widetext}
\begin{eqnarray}
S_{N}^{J}&=&\bigg|\sum_{\{x_{i}\}^{'}}\bigg[v(x_{1}=0,x_{2},...,x_{N})E(A_{1}^{J}A_{2}\cdot\cdot\cdot A_{N}|x_{1}=0,x_{2},...,x_{N})\nonumber\\
&&+v(x_{1}=1,x_{2},...,x_{N})E(A_{1}^{J'}A_{2}\cdot\cdot\cdot A_{N}|x_{1}=1,x_{2},...,x_{N})\bigg]\bigg|\nonumber\\
&=&\left|\sum_{\{x_{i}\}^{'}}v(x_{1}=0,x_{2},...,x_{N})\big[E(A_{1}^{J}A_{2}\cdot\cdot\cdot A_{N}|x_{1}=0,x_{2},...,x_{N})+(-1)^{k'}E(A_{1}^{J'}A_{2}\cdot\cdot\cdot A_{N}|x_{1}=1,x_{2},...,x_{N})\big]\right|,\nonumber\\
\label{dd2}
\end{eqnarray}
where $k'$ denotes the number of times index $1$ appears in $(x_{2},...,x_{N})$, $S_{N}^{J}$ is the $N$-particle Svetlichny's operator concerned with joint measurements, and $\{x_{i}\}^{'}$ stands for an $N-1$-tuple $x_{2},...,x_{N}$. There are $2^{N-1}$ terms in the summation $\sum_{\{x_{i}\}^{'}}$.
If the number of times index $1$ appears in $(x_{2},...,x_{N})$ is even, i.e. $k'$ is even, we can obtain
\begin{eqnarray}
&&v(x_{1}=0,x_{2},...,x_{N})\big[E(A_{1}^{J}A_{2}\cdot\cdot\cdot A_{N}|x_{1}=0,x_{2},...,x_{N})+(-1)^{k'}E(A_{1}^{J'}A_{2}\cdot\cdot\cdot A_{N}|x_{1}=1,x_{2},...,x_{N})\big]\nonumber\\
&\leq&E(A_{1}^{J}A_{2}\cdot\cdot\cdot A_{N}|x_{1}=0,x_{2},...,x_{N})+E(A_{1}^{J'}A_{2}\cdot\cdot\cdot A_{N}|x_{1}=1,x_{2},...,x_{N})\nonumber\\
&=&2\big[P(A_{1}^{J}=A_{1}^{J'}=A_{2}\cdot\cdot\cdot A_{N}|x_{2},...,x_{N})-P(A_{1}^{J}=A_{1}^{J'}=-A_{2}\cdot\cdot\cdot A_{N}|x_{2},...,x_{N})\big]\nonumber\\
&\leq&2\big[P(A_{1}^{J}=A_{1}^{J'}=A_{2}\cdot\cdot\cdot A_{N}|x_{2},...,x_{N})+P(A_{1}^{J}=A_{1}^{J'}=-A_{2}\cdot\cdot\cdot A_{N}|x_{2},...,x_{N})\big]\nonumber\\
&=&2P(A_{1}^{J}=A_{1}^{J'}|x_{2},...,x_{N}),
\label{dd3}
\end{eqnarray}
where $P(A_{1}^{J}=A_{1}^{J'}|x_{2},...,x_{N})$ represents the probability of that the joint measurement results of the first player satisfy $A_{1}^{J}=A_{1}^{J'}$ when the measurement setting of other players is $x_{2},...,x_{N}$.
Similarly, if $k'$ is odd we can obtain
\begin{eqnarray}
&&v(x_{1}=0,x_{2},...,x_{N})\big[E(A_{1}^{J}A_{2}\cdot\cdot\cdot A_{N}|x_{1}=0,x_{2},...,x_{N})+(-1)^{k'}E(A_{1}^{J'}A_{2}\cdot\cdot\cdot A_{N}|x_{1}=1,x_{2},...,x_{N})\big]\nonumber\\
&\leq&E(A_{1}^{J}A_{2}\cdot\cdot\cdot A_{N}|x_{1}=0,x_{2},...,x_{N})-E(A_{1}^{J'}A_{2}\cdot\cdot\cdot A_{N}|x_{1}=1,x_{2},...,x_{N})\nonumber\\
&=&2\big[P(A_{1}^{J}=-A_{1}^{J'}=A_{2}\cdot\cdot\cdot A_{N}|x_{2},...,x_{N})-P(A_{1}^{J}=-A_{1}^{J'}=-A_{2}\cdot\cdot\cdot A_{N}|x_{2},...,x_{N})\nonumber\\
&\leq&2\big[P(A_{1}^{J}=-A_{1}^{J'}=A_{2}\cdot\cdot\cdot A_{N}|x_{2},...,x_{N})+P(A_{1}^{J}=-A_{1}^{J'}=-A_{2}\cdot\cdot\cdot A_{N}|x_{2},...,x_{N})\nonumber\\
&=&2P(A_{1}^{J}=-A_{1}^{J'}|x_{2},...,x_{N}),
\label{dd4}
\end{eqnarray}

Due to the no-signaling principle, the probability of the first player getting $A_{1}^{J}=A_{1}^{J'}$ or $A_{1}^{J}=-A_{1}^{J'}$
should be independent of the measurement choices of other players, i.e.
\begin{eqnarray}
&&P(A_{1}^{J}=A_{1}^{J'}|x_{2},...,x_{N})=P(A_{1}^{J}=A_{1}^{J'}),~~P(A_{1}^{J}=-A_{1}^{J'}|x_{2},...,x_{N})=P(A_{1}^{J}=-A_{1}^{J'}).
\label{dd5}
\end{eqnarray}
The number of terms with even $k'$ in the summation $\sum_{\{x_{i}\}^{'}}$ in Eq. (\ref{dd2}) is $2^{N-2}$ (${N-1 \choose 0}+{N-1 \choose 2}+...=2^{N-2}$),
and there are also $2^{N-2}$ terms with odd $k'$ in the summation $\sum_{\{x_{i}\}^{'}}$.
So from Eq. (\ref{dd2}) we finally get the $N$-particle SI
\begin{eqnarray}
S_{N}^{J}&\equiv&\left|\sum_{\{x_{i}\}^{'}}v(x_{1}=0,x_{2},...,x_{N})\big[E(A_{1}^{J}A_{2}\cdot\cdot\cdot A_{N}|x_{1}=0,x_{2},...,x_{N})+(-1)^{k'}E(A_{1}^{J'}A_{2}\cdot\cdot\cdot A_{N}|x_{1}=1,x_{2},...,x_{N})\big]\right|\nonumber\\
&\leq&\left|\left[{N-1 \choose 0}+{N-1 \choose 2}+...\right]\cdot2P(A_{1}^{J}=A_{1}^{J'})+\left[{N-1 \choose 1}+{N-1 \choose 3}+...\right]\cdot2P(A_{1}^{J}=-A_{1}^{J'})\right|\nonumber\\
&=&2^{N-1}.
\label{s}
\end{eqnarray}
\end{widetext}

\section{bound on the sharpness of joint measurements and the maximal violation of SI}

If $a$ and $a'$ both denote the usual projective quantum measurements, they can be described by projector collections of
$\{E(\bf{a}), E(\bf{-a})\}$ and $\{E(\bf{a'}), E(\bf{-a'})\}$ respectively, where $E(\pm{\bf a})=\frac{1}{2}[\I \pm{\bf a} \cdot {\bf\sigma}]$, $E(\pm{\bf a'})=\frac{1}{2}[\I \pm{\bf a'} \cdot {\bf\sigma}]$, ${\bf a}$ and ${\bf a'}$ are the directions
of the measurements $a$ and $a'$, and ${\bf\sigma}$ is Pauli operator. Suppose $a$ and $a'$ do not mutually commute we cannot perform a joint measurement of
$a$ and $a'$ by doing a projective quantum measurement, since the observables $a$ and $a'$ do not share eigenstates. However
quantum mechanics allows us to perform joint unsharp measurements of these two observables, and the unsharp measurements for these
two observables can be described as POVMs . We can describe the unsharp measurements of $a$ and $a'$ respectively as \cite{and1, and2,busch}
\begin{eqnarray}
E_{\eta_{1}}(\pm{\bf a})=\frac{1}{2}[\I \pm\eta_{1}{\bf a} \cdot {\bf\sigma}],\nonumber\\
E_{\eta_{2}}(\pm{\bf a'})=\frac{1}{2}[\I \pm\eta_{2}{\bf a'} \cdot {\bf\sigma}],
\label{povm}
\end{eqnarray}
where $0<\eta_{i}\leq 1$ quantifies the sharpness of the joint measurements of $a$ and $a'$.
For a joint measurement of $\{E_{\eta_{1}}(\pm{\bf a})\}$ and $\{E_{\eta_{2}}(\pm{\bf a'})\}$, the only constraint on the ${\bf a}$, ${\bf a'}$,
$\eta_{1}$, and $\eta_{2}$ is \cite{and1,bus1}
\begin{eqnarray}
\left|\eta_{1}{\bf a}+\eta_{2}{\bf a'}\right|+\left|\eta_{1}{\bf a}-\eta_{2}{\bf a'}\right|\leq 2.
\label{condition1}
\end{eqnarray}
If we take $\eta_{1}=\eta_{2}=\eta$, which means that the measurements of $\{E_{\eta_{1}}(\pm{\bf a})\}$ and $\{E_{\eta_{2}}(\pm{\bf a'})\}$
have equal sharpness, then the above condition can be expressed as
\begin{eqnarray}
\eta\left[\left|{\bf a}+{\bf a'}\right|+\left|{\bf a}-{\bf a'}\right|\right]\leq 2.
\label{condition2}
\end{eqnarray}
For the case of ${\bf a}\perp{\bf a'}$, $\left[\left|{\bf a}+{\bf a'}\right|+\left|{\bf a}-{\bf a'}\right|\right]$ takes its maximal value of $2\sqrt{2}$,
so from Eq. (\ref{condition2}) we know that quantum mechanics allows joint unsharp measurements of any observables $a$ and $a'$ as long as
the equal sharpness $\eta$ is less than or equal to $\frac{\sqrt{2}}{2}$.

According to quantum measurement theory, there is an essential property of unsharp measurement of $E_{\eta}(\pm{\bf a})=\frac{1}{2}[\I \pm\eta{\bf a} \cdot {\bf \sigma}]$. For any state
$\rho$, the average value of the measurement results of $\{E_{\eta}(\pm{\bf a})\}$ is proportional to the expectation value of
the corresponding sharp measurement, i.e.
\begin{eqnarray}
\tr\left[\rho E_{\eta}({\bf a})\right]-\tr\left[\rho E_{\eta}(-{\bf a})\right]=\eta\tr\left[\rho{\bf a}\cdot{\bf \sigma}\right].
\label{exp1}
\end{eqnarray}

Now we assume the first player makes a joint unsharp measurement of $a_{1}$ and $a_{1}'$ on the first particle, with
equal sharpness $\eta$. From Eq. (\ref{exp1}) we can obtain $S_{N}^{J}=\eta S_{N}$. The condition of Eq. (\ref{condition2}) means that, as long as the equal sharpness $\eta$ satisfies
$\eta\leq\frac{\sqrt{2}}{2}$ quantum mechanics allows the existence of joint unsharp measurements of any observables $a$ and $a'$.
The derivation in the previous section shows that the existence of joint measurements of two observables $a_{1}$ and $a_{1}'$ must
demand that $S_{N}^{J}\leq 2^{N-1}$ or the no-signaling principle will be breached. So we can conclude that the quantum bound of $S_{N}$ is necessarily
not greater than $2^{N-1}\sqrt{2}$, and this bound can be achieved in quantum mechanics with the GHZ
state of $\frac{1}{\sqrt{2}}\left(\ket{\uparrow}^{\otimes N}_{z}\pm\ket{\downarrow}^{\otimes N}_{z}\right)$
and a proper measurement protocol \cite{seev}.

\section{conclusion}

The feasibility of joint measurements of some observables implies that there must be a joint probability distribution of the values of these observalbles, so we
can derive SI from the existence of the joint measurements and the no-signaling principle. This means that, if we do not breach the no-signaling
principle the results of joint measurements must produce the satisfaction of SI. Quantum mechanics allows joint unsharp measurement of any observables
as long as the equal sharpness fulfills $\eta\leq\frac{\sqrt{2}}{2}$, thus the quantum bound of $S_{N}$ is necessarily
not greater than $2^{N-1}\sqrt{2}$, otherwise we would get $S_{N}^{J}>2^{N-1}$ and this implies the breach of the no-signaling principle.


\section*{Acknowledgments}
~This work is supported by National Foundation of Natural Science in
China under Grant Nos. 10947142 and 11005031.

\end{document}